\documentclass[conference]{IEEEtran}\IEEEoverridecommandlockouts
\usepackage{amssymb}
\usepackage{amsmath}
\usepackage{amsfonts}
\usepackage{graphicx}
\usepackage{algorithm}
\usepackage{algorithmic}
\usepackage{cite}
\usepackage{epstopdf}
\usepackage{stfloats}
\newtheorem{remark}{\textbf{Remark}}

\begin{document}
\bibliographystyle{IEEEtran}

\title{Cache-enabled Uplink Transmission in Wireless Small Cell Networks}
%\title{Energy-Efficient Transmission over Cache-enabled Uplink Channels with Redundancy Elimination}
%\title{Uplink Cache-based Small Cell Networks with Redundancy Elimination}
\author{Zhanzhan Zhang$^{*}$, Zhiyong Chen$^{*}$, Hao Feng$^{*}$, Bin Xia$^{*}$, Weiliang Xie$^{\dag}$, and Yong Zhao$^{\dag}$\\
${^*}$Department of Electronic Engineering, Shanghai Jiao Tong University, Shanghai, P. R. China\\
$^\dag$China Telecom Corporation Limited Technology Innovation Center\\
Email: \{mingzhanzhang, zhiyongchen, fenghao, bxia\}@sjtu.edu.cn, \{xiewl, zhaoyong\}@ctbri.com.cn}

\maketitle

\begin{abstract}
It is starting to become a big trend in the era of social networking that people produce and upload user-generated contents to Internet via wireless networks, bringing a significant burden on wireless uplink networks. In this paper, we contribute to designing and theoretical understanding of wireless cache-enabled upload transmission in a delay-tolerant small cell network to relieve the burden, and then propose the corresponding scheduling policies for the small base station (SBS) under the infinite and finite cache sizes. Specifically, the cache ability introduced by SBS enables SBS to eliminate the redundancy among the upload contents from users. This strategy not only alleviates the wireless backhual traffic congestion from SBS to a macro base station (MBS), but also improves the transmission efficiency of SBS. We then investigate the scheduling schemes of SBS to offload more data traffic under caching size constraint. Moreover, two operational regions for the wireless cache-enabled upload network, namely, the \emph{delay-limited} region and the \emph{cache-limited} region, are established to reveal the fundamental tradeoff between the delay tolerance and the cache ability. Finally, numerical results are provided to demonstrate the significant performance gains of the proposed wireless cache-enabled upload network.
\end{abstract}

\section{Introduction}
With the significant growth of mobile Internet, which offers a convenient way to exchange information ubiquitously, people are no longer only consuming content but have started creating content. For example, users can capture the real-time events using smartphones and share them with other users through mobile applications, such as YouTube or Facebook. This growing trend of user-generated content (UGC) leads to the unprecedented increase of mobile data traffic and imposes a great uploading pressure to the wireless networks.

To cope with the \emph{mobile data tsunami} of mobile data traffic, lots of research efforts have been devoted towards content downloading in mobile cellular networks \cite{BUPT,content,Mobile3C}, while little efforts have been made to facilitate the uplink of mobile cellular networks to meet the increasing demands of UGC uploading. Different with the downloading, there exist many constraints for the upload of multimedia contents. First, the mobile cellular networks are asymmetric in terms of bandwidth between downlink and uplink communications. It is reported that the downlink bandwidth could be 10--1000 times the uplink bandwidth\cite{ToN15asymmetry}, yielding the low throughput on the uplink channels, long upload time and degraded quality of experience (QoE) as uploading multimedia contents. Second, the resources of mobile devices are limited, such as the caching size, the transmit power and battery capacity. Thus it is imperative and challenging to alleviate the upload traffic load of the mobile cellular network based on the traditional capacity-increasing solutions, e.g., additional uplink spectra.

Exploiting the caching resource in the network to reduce the duplicate content transmissions is a potential solution for the above wireless data challenge, independent of the limited communication resources. \cite{cha2007tube} reported that a large number of the edited videos were uploaded on the same day as the original video or within a week. Besides, contents can be modularized today. For example, the Moving Picture Experts Group (MPEG) has developed the new MPEG Media Transport (MMT) standard, with which the logical entity MMT package consists of MMT assets and information about the data combining and delivering \cite{MMT2013Mag}. In a word, the content popularity and modularization provide a higher potential to redundancy among contents by combining with caching.

There has been a few works done to address the content uploading problems with cache considered \cite{QoEdriven11C, uploadCache12C, deployableUpload12C, SOP15C}. In \cite{QoEdriven11C}, the joint upstreaming of real-time and time-shifted on-demand videos under scarce uplink resources was optimized by considering cache at mobile terminals. Then the authors in \cite{uploadCache12C} proposed a upload cache scheme in edge networks to shorten the duration and reduce the peak traffic volume, by dividing the traditional upload process into two phases: from the client to the gateway cache and from the gateway to the destination server. In addition, both \cite{deployableUpload12C} and \cite{SOP15C} also divided the upload process in the same way as in \cite{uploadCache12C} and placed cache at the nearby WiFi access points. Specifically, \cite{deployableUpload12C} demonstrated that larger contents could save more connection time for users. And \cite{SOP15C} proposed a smart offloading mechanism for content uploading and considered the WiFi bandwidth scheduling. However, how to schedule the contents in the cache and how much cache space is needed  are not addressed in \cite{uploadCache12C, deployableUpload12C, SOP15C}.

In this paper, we investigate the content uploading in a small cell network, where multiple users are served by a small base station (SBS) which connects to the core network through a macro base station (MBS). The SBS is equipped with a cache memory for temporarily buffering the received contents from users, which implies that data transfers can be delayed at the SBS with some deadlines, this delay assumption is reasonable in most cases, such as in upload cache \cite{uploadCache12C} and WiFi offloading \cite{wifiOffloading13ToN}. The SBS cache space and the content delay tolerance enable the SBS to perform redundancy elimination among similar files. Our contributions are summarized as follows:
\begin{itemize}
  \item We propose upload cache in small cell networks based on redundancy elimination, which improves the transmission efficiency of the SBS and the effective bandwidth.
  \item Scheduling policies of the SBS on cached contents are derived based on the probabilistic knowledge of users' future upload requests to further improve the efficiency.
  \item Elaborated numerical results provide valuable insights on how to design the system parameters, such as the cache size, delay tolerance and the user number.
\end{itemize}
\begin{figure}
  \centering
  \includegraphics[width=3.7in]{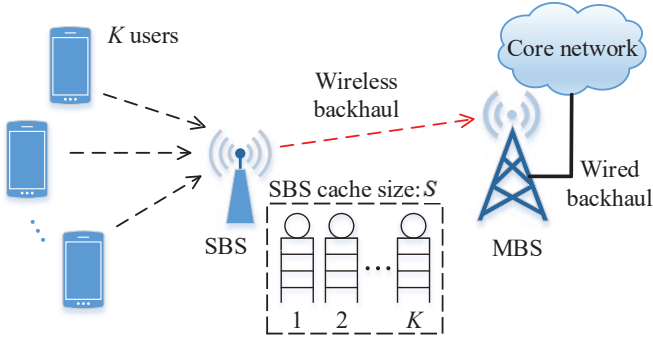}
  \caption{Uplink cache-enabled small cell networks with cache at the SBS.}\label{systemModel}
\end{figure}

\section{System Model}
\subsection{SBS Caching}
As illustrated in Fig. \ref{systemModel}, we consider the wireless uplink transmission in a cache-enabled single small cell network, where a SBS serves $K$ mobile users. In the small cell network, the SBS is connected to a MBS over a wireless backhual, and the MBS has a wired backhaul connection with the core network. Thus, one user can upload a content to Internet via the wireless transmission, wireless backhaul, and wired backhual. As a result, the wireless backhual between the SBS and the MBS is to restrict the wireless upload performance. In this paper, we propose a new architecture for wireless content uploading, where the SBS can cache some of contents from users and then do duplication elimination among the contents to reduce the backhaul payload.

In the traditional uplink processing, the user performing the content uploading will not leave the connection with the SBS until the uploading task to the target server is finished, which consumes more user's energy. In this paper, however, the
uploaded content will firstly be received and cached at the SBS temporarily, and the user will leave the system once the uploading to the SBS is finished, then the SBS starts the backhaul uploading at a proper time.

We assume that the SBS can detect and eliminate the duplication among similar contents, which requires the ability of content chunking and hashing computation, while the mobile users don't have the chunking ability due to constraints of the battery life and computing resource. Besides, the file-level deduplication technique is assumed to be employed at mobile users, i.e., one user will compute and transmit the hash value of a content before uploading. The user will finish the uploading and go on to next task if this content has already existed at the target server or at the SBS, and otherwise start the uploading.

\begin{remark}
Note that caching at the SBS has several benefits although it prolongs the overall uploading processing. First, it reduces users' online time of connecting the SBS. Second, it improves the transmission efficiency of the SBS and alleviates the backhaul congestion. Moreover, caching the most recently uploaded data at the nearest SBS can provide the user a fast review experience.
\end{remark}

\subsection{System Description}
Assume that the SBS is equipped with a finite cache memory of $S$ units. We denote $T$ as the considered time horizon which is made up of $N$ time slots and the duration of each time slot equals $T_s$. For detecting duplication among similar contents, the SBS needs to divide the received contents into a number of variable-sized chunks, e.g., MMT assets. Therefore, we assume that all the variable-sized chunk contents composing each whole content come from a set of all possible chunk files, which is denoted by $\mathbb{F}=\{ f_1, ..., f_F \}$. Chunk file $f_j$ has length of $l_j$. The maximal file length in the set $\mathbb{F}$ is denoted by $l_{max}$ and we assume that $S\geqslant K\cdot l_{max}$.
Besides, we define $p_j$ as the file popularity of chunk file $f_j$, indicating the upload probability of requesting file $f_j$.
Without loss of generality, we only consider a file in the chunk-level in the following.

The SBS allocates a virtual cache queue space for each user, as shown in Fig. \ref{systemModel}. In each time slot, each user, $k$ ($k = 1, \cdots, K$), requests to upload a file to its cache queue from the set $\mathbb{F}$ based on the file popularity. The transmitting duration of each file is fixed to be $T_s$, then file $f_j$ is transmitted at a constant rate $l_j/T_s$. In addition, a user is assumed to transmit file $f_0$ with zero length if it keeps silent in a time slot. We consider that the arisings of contents among different time slots in a cache queue are independent\footnote{This means that the same contents may arise in a cache queue in different time slots. In reality, a user is less likely to upload two same contents during a short time. But usually, there are more than $K$ users in a small cell. For a cache queue, another user will replace the current user's position if the current user leave the system after finishing its upload, and the total number of users that have data to send keeps $K$. Thus this independent assumption is reasonable.}, so are the arisings among different cache queues.

We consider a delay-tolerant network, in which we denote by $t_d$ the duration of stay that one user can tolerate its content being cached at the SBS, and one content must be transmitted by the SBS when $t_d$ expires or before. Without loss of generality, we consider $t_d = n_dT_s$ ($n_d\in \mathbb{Z}^+$). Once a time slot ends, the SBS compares the just received data with the cached contents immediately. If there is duplication, the SBS will delete the new received data and keep the corresponding earlier version to meet the deadline requirement. Note that enough auxiliary information will be created to reconstruct the whole content in the destination server\footnote{For example, the auxiliary information can be file recipes which are used to rebuild the whole files based on chunk files and their hash values. Except those, the hash values of whole files are also needed to be stored in the SBS cache temporarily as auxiliary information for file-level duplication detection.}. Meanwhile, the SBS determines which contents to be transmitted based on a scheduling policy, in order to satisfy the cache memory and the deadline requirements.

%\subsection{Definitions}
We denote by $\delta_k(j,i)$ the \textit{file upload indicator} which equals $1$ if user $k$ chooses to upload file $f_j$ ($j = 1, ..., F$) in time slot $i$ and value $0$, otherwise (upload file $f_0$). Besides, one user can only transmit one content in every time slot, as a result, we get $\sum_{j=0}^{F}\delta_k(j,i)=1$.

\vspace{0mm}

\textit{\textbf{Definition 1}} (\textit{File Arrival Rate}). We define $d_k(t)$ ($t\in[0,T]$) as the file arrival rate at which user $k$ transmits data to the SBS, and it's given by $d_k(t) = \sum_{i=1}^{N}d_{k,i}\mathrm{rect}\left( \frac{t-(i-\frac{1}{2})T_s}{T_s} \right)$,
where $d_{k,i}=\sum_{j=0}^{F}\delta_k(j,i)\frac{l_j}{T_s}$ represents the transmit rate of user $k$ in time slot $i$, and $\mathrm{rect}(\frac{t-a}{b})$ indicates the rectangular function which is centered at $a$ and has duration $b$.

\vspace{0mm}

\textit{\textbf{Definition 2}} (\textit{File Arrival Rate after Deduplication}). We define $v_k(t)$ ($t\in[0, t_{ps}]$) as the file arrival rate of user $k$ seen by the SBS after duplication elimination, and $t_{ps}=nT_s$ stands for the present scheduling time instant (STI). Note that multiple users might upload one same file simultaneously which the current cache doesn't contain. Without loss of generality, we only keep the uploaded version from the user with the smallest index $k$, and delete the other counterparts. As a result, let $\sigma_k(j,n)$ denote the \textit{file indicator after deduplication} which takes value $1$ when user $k$ is scheduled to upload file $f_j$ ($j\neq 0$) in time slot $n$ with the smallest index $k$ and no duplication is detected in the cache, and $0$ otherwise. Therefore, we have $v_k(t) = \sum_{i=1}^{n}v_{k,i}\mathrm{rect}\left( \frac{t-(i-\frac{1}{2})T_s}{T_s} \right)$,
where $v_{k,i}=\sum_{j=0}^{F}\sigma_k(j,i)\frac{l_j}{T_s}$ signifies the file arrival rate of user $k$ in time slot $i$  after deduplication.

\vspace{0mm}

\textit{\textbf{Definition 3}} (\textit{File Status Indicator}). We define $c_{k,i}^n$ ($i\leqslant n$) as the status of the file transmitted by user $k$ in the $i$-th time slot (this file is also denoted by $f_{k,i}$) when the $n$-th time slot ends. We have $c_{k,i}^n=1$ if $f_{k,i}$ is still buffered in the SBS at $t_{ps}=nT_s$, and $c_{k,i}^n=0$, otherwise. Besides, $c_{k,n}^n$ is given by $c_{k,n}^n = \sum_{j=0}^F { \sigma_k(j,n) }$.

\vspace{0mm}

\textit{\textbf{Definition 4}} (\textit{SBS Scheduling Indicator}). We define $a_{k,i}^n$ ($i\leqslant n$) as the SBS scheduling indicator at $t_{ps}=nT_s$ with regard to the file $f_{k,i}$. We have $a_{k,i}^n=1$ if file $f_{k,i}$ is going to be transmitted by the SBS in the ($n+1$)-th time slot, and $a_{k,i}^n=0$, otherwise. Therefore, we obtain $c_{k,i}^{n+1} = 1 - a_{k,i}^n$.

Accordingly, the transmit rate of the SBS in the ($n+1$)-th time slot can be expressed as $r_{n+1} = \sum_{k=1}^K \sum_{i=1}^n c_{k,i}^n a_{k,i}^n v_{k,i}$.

\section{SBS Scheduling Policy}
In this section, for reducing the total data volume that are transmitted by the SBS in the long term, we propose the SBS scheduling strategies on the cached contents. Besides, we consider the online scheduling policy, i.e., the SBS does not know what contents the users are going to upload exactly, but only knows the probabilistic knowledge of users' future upload requests. Therefore, the SBS cannot make scheduling policies for the future time slots just like the offline scenario \cite{GregoriJSAC16} where the complete knowledge of the future upload requests is known. The SBS can only make scheduling decisions at every time slot transition, under the constraints of the contents deadlines and the limited SBS cache.

We denote by $D_0$ the data volume transmitted by the SBS to finish the upload requests in $N$ time slots if no cache and no deduplication technique are employed, and by $D_1$ the corresponding data volume when cache and the deduplication technique are adopted. Then, we have
%\begin{equation}\label{D0}
%D_0 = \sum_{i=1}^{N}\sum_{k=1}^{K} d_{k,i} T_s,
%\end{equation}
%\begin{equation}\label{D1}
%D_1 = \sum_{i=1}^{N}\sum_{k=1}^{K} v_{k,i} T_s.
%\end{equation}
\begin{align}
&D_0 = \sum_{i=1}^{N}\sum_{k=1}^{K} d_{k,i} T_s, \label{D0}\\
&D_1 = \sum_{i=1}^{N}\sum_{k=1}^{K} v_{k,i} T_s. \label{D1}
\end{align}
We consider the percentage of the saved data traffic that are eliminated due to duplication  as
the performance metric of the system, which is denoted by $\eta = \frac{D_0-D_1}{D_0} \cdot 100\%$.

For disclosing the insights of the impact of the cache memory size on the system performance, we consider two cases: 1) infinite cache and 2) finite cache. And for simplicity, the deadlines for all contents in the set $\mathbb{F}$ are assumed to be the same.
\subsection{Infinite SBS Cache}
When $S\rightarrow \infty$, the SBS makes scheduling decisions only based on the content deadlines, $t_d$. In order to eliminate as much duplicated data as possible, each content will always stay in the SBS cache to be used for duplication detection unless it has to be scheduled. As a result, the SBS just schedules to transmit all the contents whose deadlines expire at every STI, when the cache size is infinite.  Thus this case is also called the \textbf{\emph{delay-limited region}}, where increasing the contents deadlines can improve the uploading performance.

\vspace{0mm}
\subsection{Finite SBS Cache}
%\subsubsection{Online Policy}
When the SBS cache size is finite, the scheduling strategies of the SBS are dependent on the cache utilization, the deadlines of cached contents and the probabilistic knowledge of users' future upload requests. In addition, we assume that the SBS is able to access the users' upload rates in the following time slot ($d_{k,n+1}$) at $t_{ps}$, which helps to ensure the unused cache is enough to hold the upcoming contents.

We denote by $S_u^n$ the used cache space after duplication elimination at $t_{ps}$.
%\begin{equation}\label{usedCacheSpace1}
%S_u^n = \sum_{k=1}^K \sum_{i=1}^n c_{k,i}^n \sum_{j=0}^F \sigma_k(j,i) l_j.
%\end{equation}
For content $f_{k,i}$ that $c_{k,i}^n=1$ at $t_{ps}$, we denote by $w_{k,i}$ and $p_{k,i}$ the corresponding file length and the probability that it is uploaded by one user in a future time slot, respectively. Thus we have $w_{k,i} = v_{k,i}T_s$ and $p_{k,i} = \sum_{j=0}^F \sigma_k(j,i) p_j$.
%\begin{equation}\label{fileLength1}
%w_{k,i} = v_{k,i}T_s = \sum_{j=0}^F \sigma_k(j,i)l_j,
%\end{equation}
%\begin{equation}\label{prob1}
%p_{k,i} = \sum_{j=0}^F \sigma_k(j,i) p_j.
%\end{equation}

Note that the contents transmitted at time slot $i_n=n-n_d+1$ will expire at $t_{ps}$ and have to be transmitted by the SBS at $t_{ps}$.
As a result, when the SBS cache size is finite, the SBS has to make scheduling among the unexpired cached contents at $t_{ps}$ in the following scenario:
\begin{itemize}
  \item when $\sum_{k=1}^K d_{k,n+1} T_s > S - S_u^n + \sum_{k=1}^K c_{k,i_n}^n w_{k,i_n}$,
which means there is no enough cache space to contain the upcoming upload contents.
\end{itemize}

Now, there is a problem about \emph{how to decide what contents to be transmitted and what to stay}. Thus we define the cache benefit score (CBS) for each content buffered in the cache in the following.

The remaining time to deadline for file $f_{k,i}$ is given by $t_i^n = \left[ n_d - ( n-i+1 ) \right] T_s$.
Considering that there are $K$ users, thus the maximal times that file $f_{k,i}$ may be transmitted in the following $t_i^n/T_s$ time slots is $q_i^n = K\cdot [ n_d - ( n-i+1 ) ]$.
Then the probability that file $f_{k,i}$ will be transmitted at least once in the following $t_i^n/T_s$ time slots is denoted by $p_{k,i}^n = 1-\left(1-p_{k,i}\right)^{q_i^n}$.
Note that file $f_{k,i}$ must be transmitted by the SBS at time instant ($t_{ps} + t_i^n$) or before, thus its CBS doesn't involve the time slots after ($t_{ps} + t_i^n$).

\textit{\textbf{Definition 5}} (\textit{Cache Benefit Score}).
We define CBS of file $f_{k,i}$ ($c_{k,i}^n=1$) at $t_{ps}$ as the difference between the probable data volume to be deleted if  $f_{k,i}$ will not be transmitted until its deadline and that if file $f_{k,i}$ will be transmitted immediately at $t_{ps}$, and it's given by
\begin{equation}\label{CBSaUser1}
\mathrm{CBS}(f_{k,i}|c_{k,i}^n=1) =  \left[ 1-\left(1-p_{k,i}\right)^{q_i^n} \right] w_{k,i}.
\end{equation}

To detect more duplication with the remaining contents, the scheduling policy tries to maximize the CBS of the contents which are chosen to stay. Let $b_{k,i}^n = 1 - a_{k,i}^n$, thus the value of $b_{k,i}^n$ has the reverse meaning of $a_{k,i}^n$, i.e., file $f_{k,i}$ will stay in the SBS cache if $b_{k,i}^n=1$, or be transmitted if $b_{k,i}^n=0$.
In addition, we have $r_{n+1}T_s \leqslant S_u^n$ since $S \geqslant Kl_{max}$, which implies that buffer overflow will not be triggered.

Then, the scheduling problem can be formulated by the following maximization problem
\begin{align}
& \max_{b_{k,i}^n} ~~ \sum_{k=1}^K \sum_{i=i_0}^n c_{k,i}^n b_{k,i}^n \left[ 1-\left(1-p_{k,i}\right)^{q_i^n} \right] w_{k,i}, \tag{4a} \label{miniProbObj1} \\
& \hspace{1.5mm} \mathrm{s.t.} \hspace{4mm} \sum_{k=1}^K \sum_{i=i_0}^n c_{k,i}^n b_{k,i}^n w_{k,i} \leqslant S -  \sum_{k=1}^K d_{k,n+1}T_s, \tag{4b} \label{miniProbSub1}\\
&\hspace{11mm} \forall c_{k,i}^n=1,  \tag{4c}
\end{align}
where $i_0=\max\{1,i_n+1\}$, and (\ref{miniProbSub1}) indicates the cache constraint.

Fortunately, the optimization problem is a 0-1 knapsack problem, where $\Theta^n =  \{ f_{k,i}|c_{k,i}^n=1, \forall k, i_0 \leqslant i \leqslant n \}$ is the given set of items with cardinal number $M^n=|\Theta^n|$, each item has a weight $w_{k,i}$ and a value $p_{k,i}^nw_{k,i}$, and the knapsack capacity is denoted by $C^n = S -  \sum_{k=1}^K d_{k,n+1}T_s$. It is known that the 0-1 knapsack problem is NP-hard and there are many approaches to solve this extensively-studied problem so far.
Of those, the dynamic programming (DP) is one effective and accurate way to seek the optimal solution. However, the pseudo-polynomial time complexity $O(M^nC^n)$ of DP goes very large when $M^n$ and/or $C^n$ increase. In addition, the greedy algorithm is an approximation method with which the solution may not be optimal, but it has a lower time complexity of $O(M^n\log(M^n))$. In the greedy algorithm, we can first sort the items in the set $\Theta^n$ in descending order of value per unit of weight, which is denoted by $u_j^n = p_{k,i}^n$ with the subscript index $j$ corresponding to the subscript index $(k,i)$, and then we have $u_1^n \geqslant u_2^n \geqslant \cdots \geqslant u_{M^n}^n$. Next, starting with $u_j^n$ in the descending order, the corresponding $b_j^n$ (i.e., $b_{k,i}^n$) will be taken to be $1$ if the weight  $w_j$ (i.e., $w_{k,i}$) is no larger than the remaining knapsack capacity, or $0$, otherwise. Thus we have

\setcounter{equation}{4}
\begin{equation}\label{greedySolu1}
b_{k,i}^n =b_j^n =
\begin{cases}
1, & \mathrm{if} ~ w_{j} \leqslant C^n - \sum_{m=1}^{j-1}b_{m}^n w_m, \\
0, & \mathrm{otherwise}.
\end{cases}
\end{equation}

With respect to the accuracy, the Lemma 1 in \cite{HaofengICC15} indicates that the greedy algorithm is near-optimal when the knapsack capacity ($C^n$) is relatively large compared to the average file length.

%\subsubsection{Offline Policy}
%In addition to the online scheduling policy, we also consider the
%offline scenario \cite{GregoriJSAC16} where the complete knowledge of the future upload requests is known.

%\begin{figure}
%\centering
%\includegraphics[width=3.3in,height=2.5in]{f1_DPGyDatavsCS_K5nd1_5_10_20_p.eps}
%\caption{The percentage of saved data traffic vs. the SBS cache size.}\label{DatavsCS}
%\end{figure}
%\begin{remark}
%When defining the CBS of file $f_{k,i}$, we assume that this file won't be uploaded until its deadline. However, due to the limited cache space, this file may be uploaded ahead of its deadline if some file with higher CBS arises in the cache. Therefore, in some sense, problem (4) is built based on the greedy strategy and the solution may not be optimal.
%\end{remark}
%In addition, if we only consider the following one time slot when defining CBS, then the corresponding objective function in problem (13) is given by
%\begin{equation}\label{secondObjFunc1}
%\max_{b_{k,i}^n, \forall c_{k,i}^n=1} ~~ \sum_{k=1}^K \sum_{i=i_0}^n c_{k,i}^n b_{k,i}^n \left[ 1-\left(1-p_{k,i}\right)^{K} \right] w_{k,i}.
%\end{equation}

\section{Numerical Results}
In this section, numerical results are illustrated to validate the performance gain of the proposed scheme. We consider the duration of each time slot $T_s = 10$ (s) and $F=1000$ chunk files. Besides, we consider a uniform distribution with an interval $[1,20]$ Mbits concerning the file length. And the file popularity $p_j$ obeys the Zipf distribution with skewness parameter $\alpha$ in the simulation\cite{ cha2007tube,FengZ}, and is independent and identically distributed across different time slots and different users.
%When all files are sorted in descending order of popularity, we have
%\begin{equation}\label{filePopu1}
%p_j = \frac{1/j^\alpha}{ \sum_{i=1}^F 1/i^\alpha },
%\end{equation}
The file popularity follows a uniform distribution when $\alpha=0$ and gets more skewed when $\alpha$ increases.
In addition, we set $N=20$, $K=5$ and $\alpha=1$ unless otherwise specified.

Since the scheduling policy is solved based on the future upload request probability, the numerical results are obtained by taking an average on 200 times simulations.

\vspace{0mm}

\begin{figure*}
  \centering
  \begin{minipage}[htb]{.24\linewidth}
    \includegraphics[width=1.9in,height=1.7in]{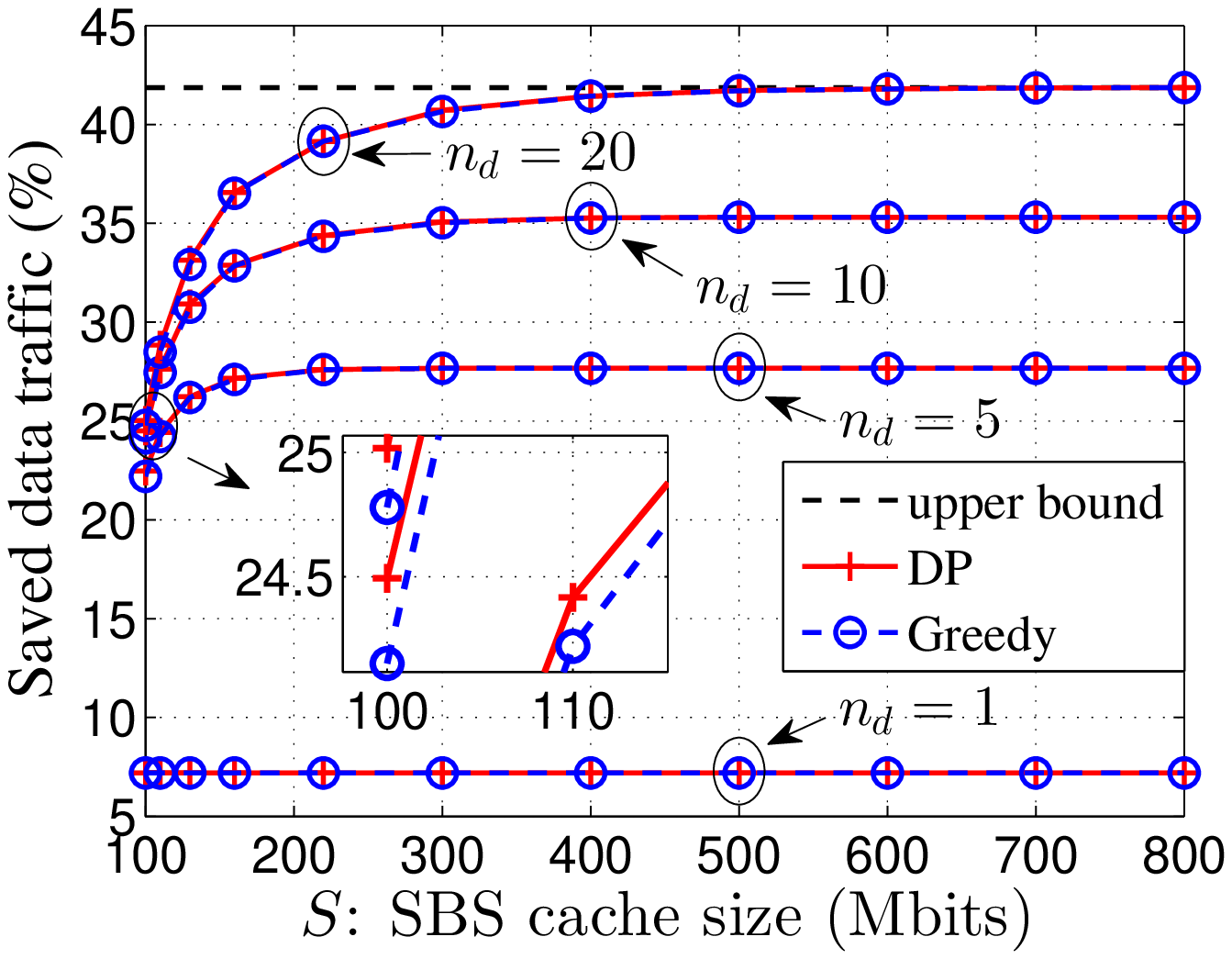}
    \caption{Saved data traffic vs. $S$.}\label{DatavsCS}
  \end{minipage}
  \begin{minipage}[htb]{.24\linewidth}
    \includegraphics[width=1.9in,height=1.7in]{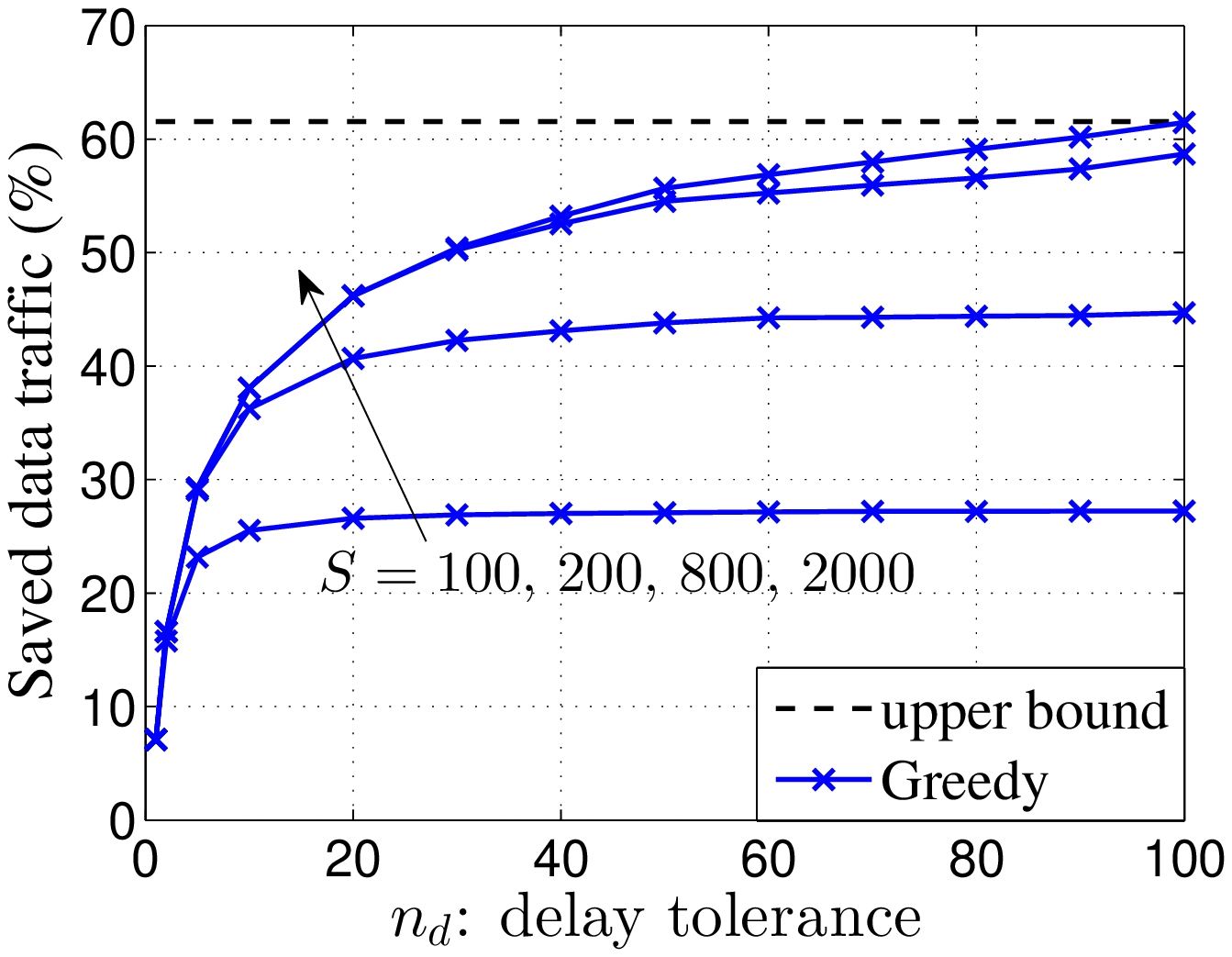}
    \caption{Saved data traffic vs. $n_d$.}\label{DatavsDelay}
  \end{minipage}
  \begin{minipage}[htb]{.25\linewidth}
    \includegraphics[width=1.9in,height=1.7in]{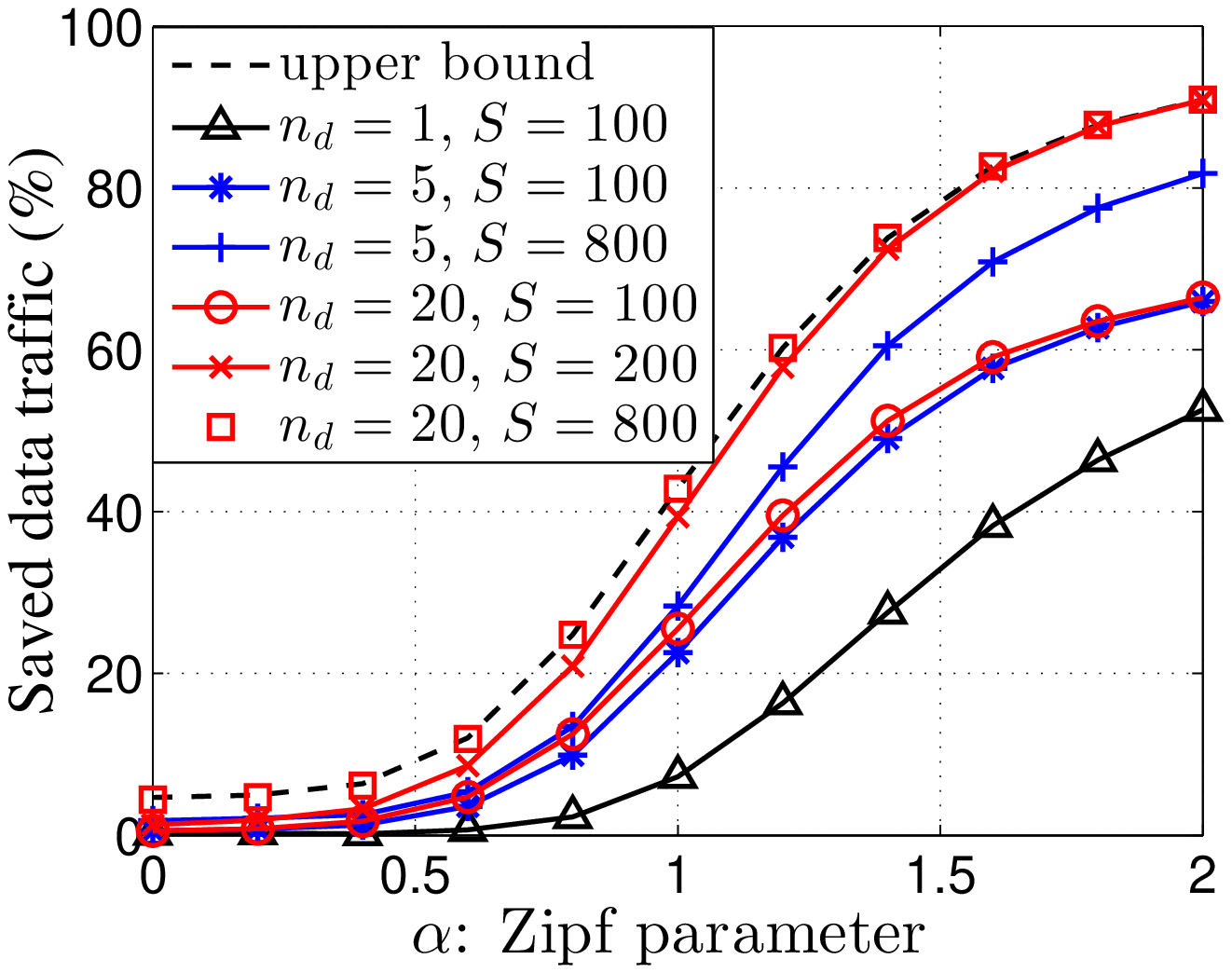}
    \caption{Saved data traffic vs. $\alpha$.}\label{DatavsAlpha}
  \end{minipage}
  \begin{minipage}[htb]{.25\linewidth}
    \includegraphics[width=1.9in,height=1.7in]{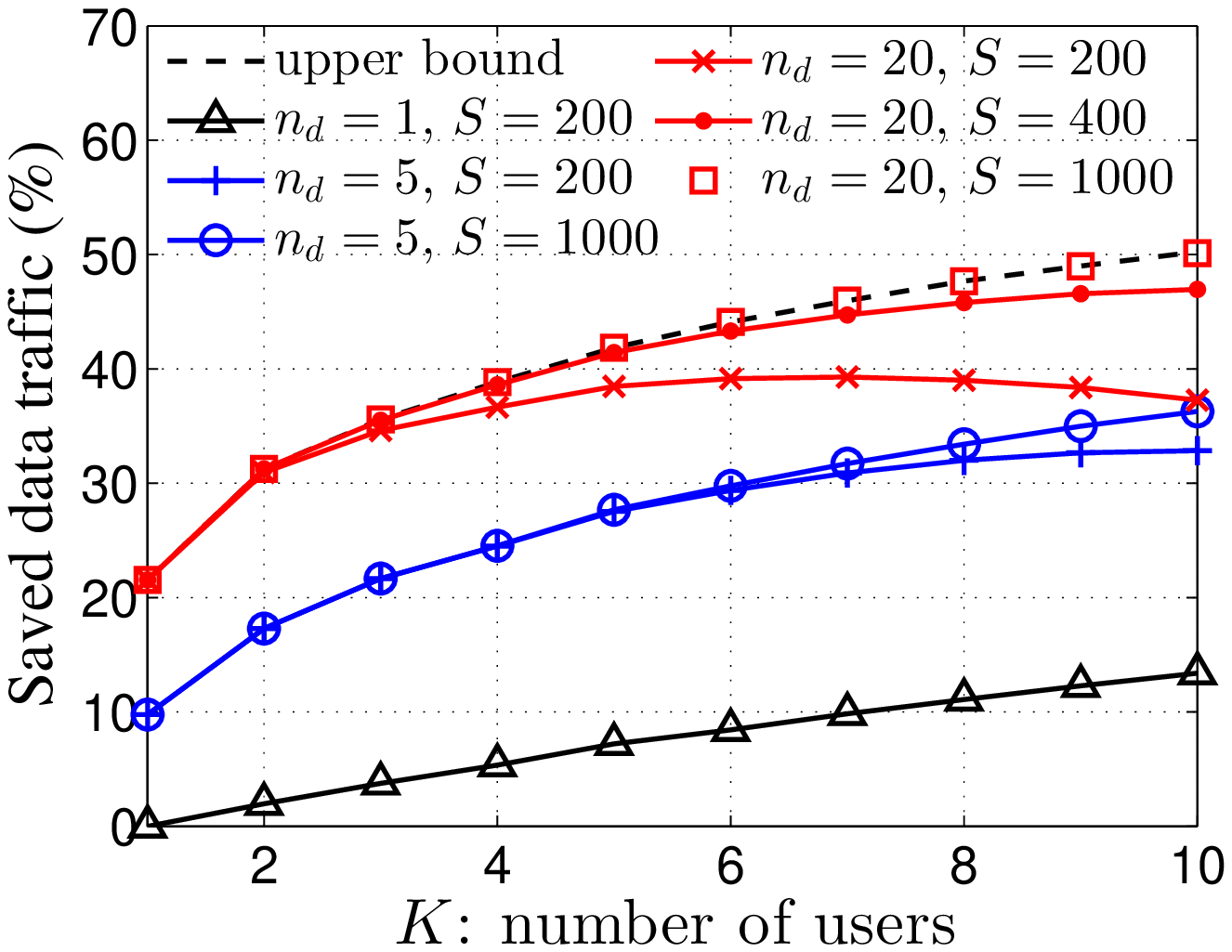}
    \caption{Saved data traffic vs. $K$.}\label{DatavsK}
  \end{minipage}
\end{figure*}

\subsection{Impact of the Cache Size and Delay Tolerance}
In Fig. \ref{DatavsCS}, the percentages of saved data traffics vs. the SBS cache size under different delay tolerances are described.  Both the DP and the greedy algorithm are used to solve knapsack problems. The upper bound is obtained by performing $\eta^{max} = (D_0-\min(D_1))/D_0$. This figure shows that the greedy algorithm suffers little performance degradation.

In Fig. \ref{DatavsCS}, we observe that the SBS can save about 7.19\% data traffic when $n_d=1$, while increasing the cache size can't bring any benefit, since cache size $S=100$ is already enough for $n_d=1$ and this corresponds to the delay-limited region. In addition, when the cache size is small (such as $S<300$), the system performance gets improved with the increase of $S$ as $n_d=5,10,20$, since small cache size will induce the 0-1 knapsack scheduling problem and the knapsack capacity increases with the cache size, thus more data can be chosen to stay in the SBS for deduplication. Moreover, enough delay tolerance and cache size (such as $n_d=20, S=800$) can reach the upper bound and save about 41.86\% data traffic.

Fig. \ref{DatavsDelay} depicts the percentages of saved data traffics vs. the delay tolerance under different SBS cache sizes considering $N=100$ time slots.
We can see that increasing the delay tolerance can achieve better performance, while in cases of $S=100$ and $S=200$, the performances will saturate when content deadlines go larger than some certain values (which grow with the increase of $S$), since increasing content deadlines doesn't change the scheduling results of the knapsack problem and these scenarios correspond to \textbf{cache-limited} region. Besides, we observe the delay-limited region (such as  $n_d\leqslant 5$ for $S\geqslant 200$ and $n_d\leqslant 30$ for $S\geqslant 800$). Furthermore, when $n_d$ falls in the middle range (such as $5\leqslant n_d \leqslant60$ for $S=200$), the performance improvement benefits from the scheduling gain of the 0-1 knapsack problem.
\vspace{0mm}

\subsection{Impact of Zipf Parameter $\alpha$}
We describe the impact of the parameter $\alpha$ of Zipf distribution on the system performance in Fig. \ref{DatavsAlpha}. It is seen that the larger the parameter $\alpha$ gets, the more data traffic the SBS can save, and the system will achieve the better performance. In addition, we observe that when $n_d=20$ and $S=200$, larger parameter $\alpha$ achieves almost the optimal performance, since we obtain the online scheduling policy based on the probabilistic knowledge of future upload requests, and the probabilities for a small number of contents will become greater as $\alpha$ grows, thus more of the contents that are scheduled to stay will match with the future upload requests and more data traffic can be saved.
\vspace{0mm}

\subsection{Impact of the Number of Users $K$}
Fig. \ref{DatavsK} describes the percentage of saved data traffic vs. the number of users. We see that in most cases, the percentage of saved data traffic grows with the increase of $K$. Specifically, it grows linearly as the number of users increases in the scenario of $n_d=1$. Besides, when $n_d=20$, $S=200$ and $K\geqslant 7$, the system suffers a little performance degradation as $K$ increases since in the cache-limited region, with larger $K$, the cached contents that the SBS can hold with the same cache size will involve less time span, then a smaller portion of duplication can be detected.
This performance degradation can be made up by adding the cache space.

\section{Conclusion}
In this paper, we proposed a upload cache system where a SBS equipped with a cache space helped the users to upload contents to servers. The contents were assumed to be delayed a certain duration of time at the SBS, which was exploited to perform the duplication elimination among similar contents. In order to improve the transmission efficiency of the SBS, scheduling policies were investigated for the cases of delay-limited region and cache-limited region, respectively. In particular, a 0-1 knapsack problem was derived for the case of cache-limited region and was efficiently solved through the greedy algorithm. The numerical results provide valuable insights on how to design the system parameters.

\appendices
%\section*{appendix}

\bibliography{reference}

\begin{thebibliography}{10}
\providecommand{\url}[1]{#1}
\csname url@samestyle\endcsname
\providecommand{\newblock}{\relax}
\providecommand{\bibinfo}[2]{#2}
\providecommand{\BIBentrySTDinterwordspacing}{\spaceskip=0pt\relax}
\providecommand{\BIBentryALTinterwordstretchfactor}{4}
\providecommand{\BIBentryALTinterwordspacing}{\spaceskip=\fontdimen2\font plus
\BIBentryALTinterwordstretchfactor\fontdimen3\font minus
  \fontdimen4\font\relax}
\providecommand{\BIBforeignlanguage}[2]{{%
\expandafter\ifx\csname l@#1\endcsname\relax
\typeout{** WARNING: IEEEtran.bst: No hyphenation pattern has been}%
\typeout{** loaded for the language `#1'. Using the pattern for}%
\typeout{** the default language instead.}%
\else
\language=\csname l@#1\endcsname
\fi
#2}}
\providecommand{\BIBdecl}{\relax}
\BIBdecl

\bibitem{BUPT}
Z.~Zhao, M.~Peng, Z.~Ding, W.~Wang, and H.~V. Poor, ``Cluster content caching:
  An energy-efficient approach to improve quality of service in cloud radio
  access networks,'' \emph{IEEE Journal on Selected Areas in Communications},
  vol.~34, no.~5, pp. 1207--1221, May 2016.

\bibitem{content}
H.~Liu, Z.~Chen, X.~Tian, X.~Wang, and M.~Tao, ``On content-centric wireless
  delivery networks,'' \emph{IEEE Wireless Commun.}, vol.~21, no.~6, pp.
  118--125, December 2014.

\bibitem{Mobile3C}
H.~Liu, Z.~Chen, and L.~Qian, ``The three primary colors of mobile systems,''
  \emph{IEEE Commun. Mag.}, vol.~54, no.~9, pp. 15--21, Sep. 2016.

\bibitem{ToN15asymmetry}
Y.~S. Li, T.~M. Cao, S.~T. Wang, and \textit{et~al.}, ``{A Resource-Constrained
  Asymmetric Redundancy Elimination Algorithm},'' \emph{IEEE/ACM Trans. Netw.},
  vol.~23, no.~4, pp. 1135--1148, Aug. 2015.

\bibitem{cha2007tube}
M.~Cha and \textit{et~al.}, ``I tube, you tube, everybody tubes: analyzing the
  world's largest user generated content video system,'' in \emph{Proc. ACM
  SIGCOMM conf. Internet measurement}, Oct. 2007, pp. 1--14.

\bibitem{MMT2013Mag}
Y.~Lim and \textit{et al.}, ``{MMT: An Emerging MPEG Standard for Multimedia
  Delivery over the Internet},'' \emph{IEEE MultiMedia}, vol.~20, no.~1, pp.
  80--85, Jan. 2013.

\bibitem{QoEdriven11C}
A.~E. Essaili, L.~Zhou, D.~Schroeder, and \textit{et~al.}, ``Qoe-driven live
  and on-demand lte uplink video transmission,'' in \emph{Proc. IEEE Int.
  Workshop Multimedia Signal Process. (MMSP)}, Oct. 2011, pp. 1--6.

\bibitem{uploadCache12C}
Y.~Zhu and A.~Nakao, ``Upload cache in edge networks,'' in \emph{Proc. IEEE
  Int. Conf. Advanced Inf. Netw. Applications (AINA)}, March 2012, pp.
  307--313.

\bibitem{deployableUpload12C}
Y.~Pu and A.~Nakao, ``A deployable upload acceleration service for mobile
  devices,'' in \emph{Proc. IEEE Int. Conf. Inf. Netw.}, Feb. 2012, pp.
  350--353.

\bibitem{SOP15C}
H.~T. Tai and \textit{et~al.}, ``Sop: Smart offloading proxy service for
  wireless content uploading over crowd events,'' in \emph{Proc. Int. Conf.
  Advanced Commun. Technol. (ICACT)}, July 2015, pp. 659--662.

\bibitem{wifiOffloading13ToN}
K.~Lee and \textit{et al.}, ``{Mobile Data Offloading: How Much Can WiFi
  Deliver?}'' \emph{IEEE/ACM Trans. Netw.}, vol.~21, no.~2, pp. 536--550, April
  2013.

\bibitem{GregoriJSAC16}
M.~Gregori, J.~G¨®mez-Vilardeb¨®, J.~Matamoros, and D.~G¨¹nd¨¹z, ``{Wireless
  Content Caching for Small Cell and D2D Networks},'' \emph{IEEE J. Sel. Areas
  Commun.}, vol.~34, no.~5, pp. 1222--1234, May 2016.

\bibitem{HaofengICC15}
H.~Feng, Z.~Chen, and H.~Liu, ``On the push-based converged network with
  limited storage,'' in \emph{Proc. IEEE Int. Conf. Commun. (ICC)}, June 2015,
  pp. 4474--4479.

\bibitem{FengZ}
------, ``Performance analysis of push-based converged networks with limited
  storage,'' \emph{IEEE Transactions on Wireless Communications}, vol.~PP,
  no.~99, pp. 1--1, 2016.

\end{thebibliography}
\end{document}